# Development of "Active Correlation" Technique

*Yu.S.Tsyganov*

*FLNR, JINR 141980 Dubna, Moscow region, Joliot-Curie 6, Russian Federation*

tyura@sungns.jinr.ru

*Abstract*

*With reaching to extremely high intensities of heavy-ion beams new requirements for the detection system of the Dubna Gas-Filled Recoil Separator (DGFRS) will definitely be set. One of the challenges is how to apply the "active correlations" method to suppress beam associated background products without significant losses in the whole long-term experiment efficiency value. Different scenarios and equations to develop the method according this requirement are under consideration in the present paper. The execution time to estimate the dead time parameter associated with the optimal choice of the life-time parameter is presented.*

1. **Introduction**

   Significant success has recently been achieved in the field of SHE synthesis and studies of radioactive properties of superheavy nuclei. With the discovery of the "island of stability" [1] in experiments with $^{48}$Ca projectiles at the Dubna Gas-Filled Recoil Separator, one can raise a question about sources and components of such a great event. Intense heavy-ion beams and exotic actinide target materials were certainly strongly required in experiments. However, final products of the DGFRS experiments were rare sequences of decaying nuclei signals. In this connection, **the** role of the DGFRS detection systems was crucial. Specifics of the DGFRS detection system is application of the "active correlations" method [2–4]. Using this technique, it has become possible to provide deep suppression of background products with negligible losses in the value of **the** whole experimental efficiency. Moreover, experiments at the DGFRS, when the above-mentioned method was not applied, yielded ambiguous results [5,6]. To briefly clarify method application, **a** process block diagram is shown in Figure 1. **A** short beam stop was generated by the EVR-α sequence detected in real-time mode. An extra time which is required for one cycle searching is shown in the Fig.1b ( i3-2100 CPU@3.10 GHz).

   Note that in most of the DGFRS experiments one of the two first alpha particle signals was used as a trigger signal for a break point in target irradiation.

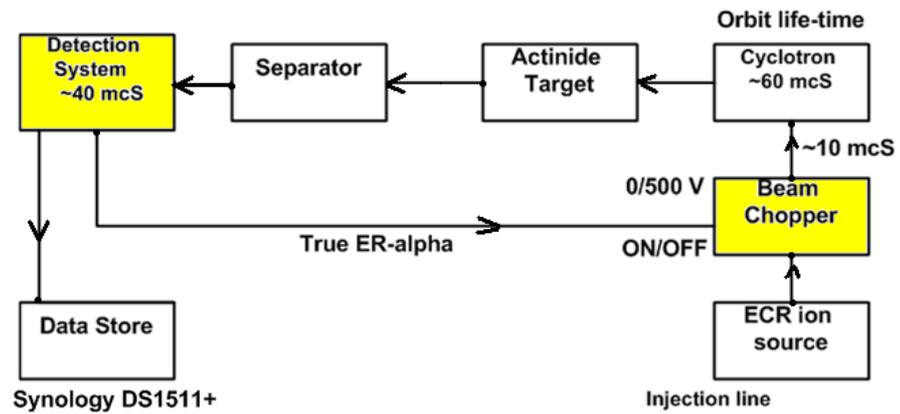

Fig. 1a. Block diagram of real-time process. Two key elements are shown in yellow color.

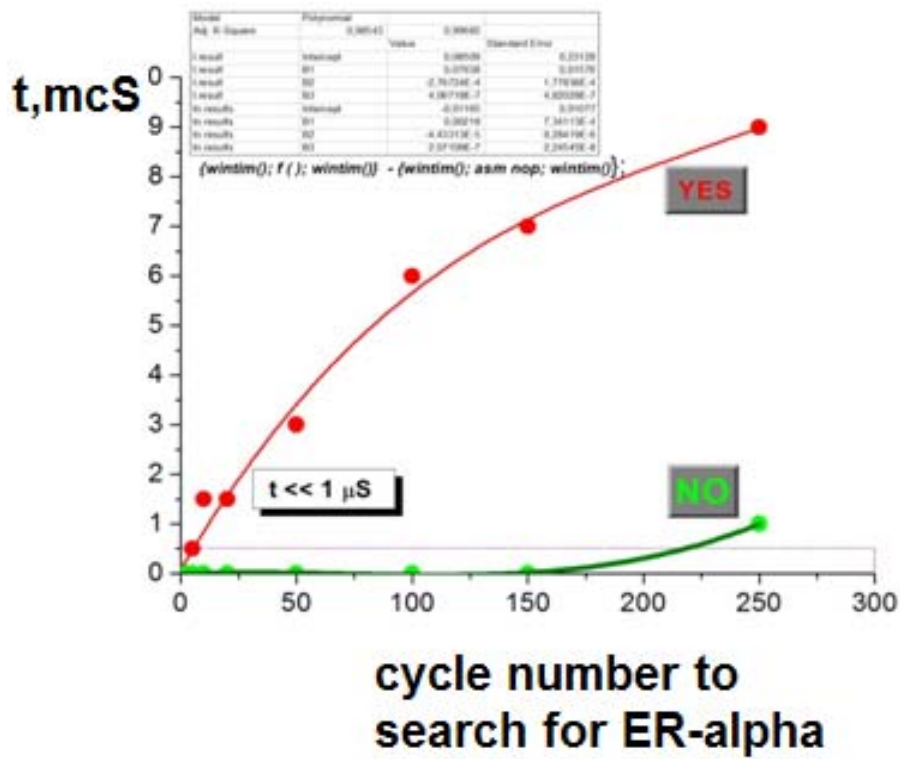

Fig.1 b Extra time t << 1 µS which is required (n=1 is actual) to search for ER-alpha correlation (scenario YES)

It is evident that application of the "active correlations" method will cause more than usual break points in continuous, long-term actinide target irradiation and, therefore, will result in additional losses in the experimental efficiency. With no any modification of **the** method, the loss value may reach tens of percents (~ 5-10 pµA intensity), which does not definitely contribute to success in challenging experiments which require **a** lot of resources. One should note that there are other problems related to high-beam intensities, except for detecting. For instance, the development of the rotating actinide target design is one of them. Additionally, this paper completes a series of works aimed to solve issues related to the DGFRS long-term experiment automation.

## 2. Two-matrix algorithm for real-time search for $EVR_1(…EVR_n)$-α sequences

It will be quite probable with extra high HI beam intensities, sequences like $EVR_1$-(…$EVR_n$-…)-α will sometimes be quite probable, though with a lower probability value than one for a single EVR-α sequence. Under these circumstances, it seems reasonable in addition to the standard algorithm [2,3] to consider two or even more candidates for a starting recoil and thus construct some virtual "effective recoil" in the PC RAM and to consider it's time as a registered alpha-decay life time value. To take into account the above-mentioned scenario, two or even $\forall$ n: n >1 EVR's matrixes in the PC's RAM are strongly required, except for one matrix in the case of the standard algorithm. In Figure 3, the flowchart of this process is shown.

In chapter 3, a general algorithm for the "virtual" recoil signal constructing is presented.

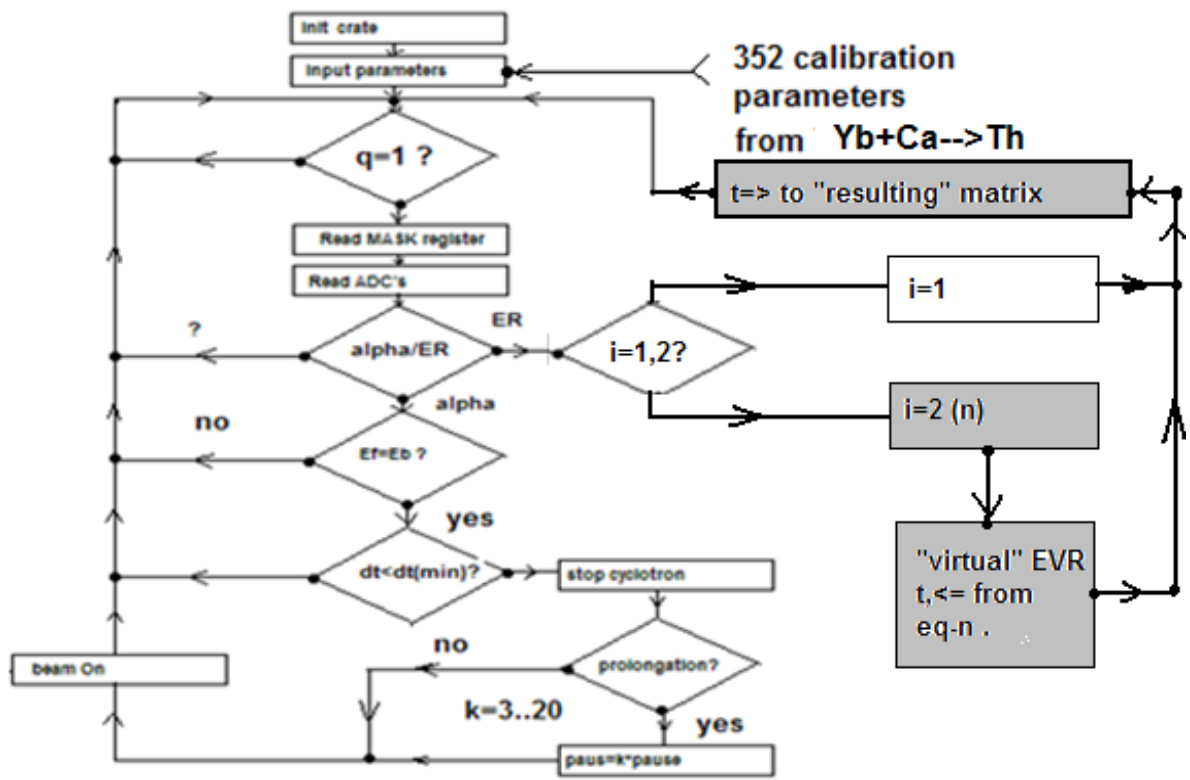

Fig.3. The flowchart of the modified process. Branch n >1 is in grey color.

### 3. *Method variation for the case of non-definite relation EVR-α*

It was V.B. Zlokazov who first recognized the importance of the theoretical approach to the non-definite mother–daughter nuclei relationship. He epitomized mathematically an equation system to search for an actual life time value [9].

A more simplified mathematical approach for two candidates for recoil (EVR) was reported in [10] in the form of a transcendental equation relative to the actual life time parameter. That equation is presented below:

$$\tau = \frac{t_1(1-e^{-\frac{t_1}{\tau}})+t_2(1-e^{-\frac{t_2}{\tau}})}{2-e^{-\frac{t_1}{\tau}}-e^{-\frac{t_2}{\tau}}}. \quad (1)$$

Here $\tau$ is the actual life time value, $t_1$ and $t_2$ are the measured times between the alpha decay and first and second recoil, respectively. In paper [10], using a simple iteration method, a 8μS time interval was found for 15 iterations, whereas additionally it was established that about a 3 μS (~3 iterations; i3-2100 CPU @ 3.10 GHz) time interval satisfied the condition of application of the real-time algorithm aimed at radical suppression of background products associated with the U-400 cyclotron beam.

Nevertheless, one may consider different values for statistical weights of different EVR signals, and sometimes it is useful to extend the above-mentioned equations to **a** more common case.

a) Let us consider n candidates for the recoil signal and time sequence $t_1$, $t_2$, $t_3$,...$t_n$, respectively (Fig.4). Of course, similar to [10], the $t_{n+1}$ signal is as follows: $t_{n+1} \gg t_i$, $i \leq n$.

Following the algorithm described in [10], it is easy to obtain:

$$\tau = \frac{\sum_{i=1}^{n} t_i(1-e^{-\frac{t_i}{\tau}})}{n-\sum_{i=1}^{n} e^{\frac{-t_i}{\tau}}}. \quad (2)$$

And with the one additional condition[i]: $1 - e^{-\frac{t_n}{<\tau_{EVR}>}} \ll 1$ . (2')

In this formula $<\tau_{EVR}>$ is an average time value between two neighbor recoil signals per a pixel. To some extent equations (2) and (2') one can consider as a contradiction. To this it is reasonable to consider (2') as a rough indication to the field of equation (2) application if to rewrite (2') in the form of:

$$1 - e^{-\frac{t_n}{\langle \tau_{EVR} \rangle}} \approx \varepsilon \text{ , where } 0 < \varepsilon < 1^{\text{ii}}.$$

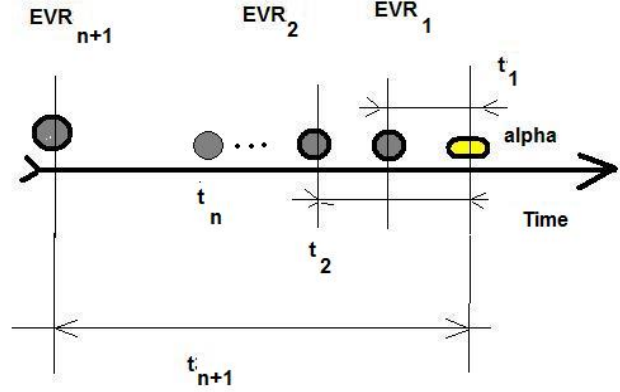

Fig.4 Schematics for the EVR(1)-EVR(2)-…-EVR(n) →α sequence. $t_{n+1} \gg t_n$.

**b)** More exactly to consider a statistical weight parameter taking into account a factor indicating to a pair of EVR's to be a random.

That is, 
$$w_i = \frac{1 - e^{-\frac{t_i}{\tau}}}{1 - e^{-\frac{t_i}{\langle \tau_{EVR} \rangle}}}. \quad (3)$$

**c)** In the approaches presented above, including [10], similar recoil signals were considered. No difference/divergence was found in energy signal amplitudes registered with the silicon radiation detector. On the contrary, a semi-empirical relationship for the EVR's registered energy value was reported in [11] in the form of:

$$< E_{REG} > \approx -2.05 + 0.73 \cdot E_{in} + 0.0015 \cdot E_{in}^2 - \left(\frac{E_{in}}{40}\right)^3. \quad (4)^{\text{iii}}$$

In this equation, $E_{REG}$ is the value of the EVR's signal registered with the PIPS (or DSSSD) silicon radiation detector and $E_{in}$ is the incoming (calculated) EVR's energy value. Taking into consideration the objective of this paper , any deviation from the mean value from **(4)** for i-recoil ( $\forall\, i \leq n$) can be introduced additionally to the $w_i$ value using the standard deviation parameter of the Gaussian distribution shape, like is reported in [12].

The time-of-flight signal value can certainly be considered in the same manner, too. *ΔE$_{start/stop}$* signals and their distribution (from "start" and "stop" proportional chambers [13, 14]) may also be taken into account. To a first approximation, due to

TOF/ΔE spectrum distribution is quite wide; one should consider a step-like function for this purpose, namely:

$$F(TOF, \Delta E) = 1: TOF \in (TOF_{MIN}, TOF_{max}) \& \Delta E \in (\Delta E_{MIN}, \Delta E_{MAX}),$$

$$0 - all\ other\ cases.\ ^{iv}(5).$$ Typical shape of ΔE signal spectrum is shown in Figure 5a. Gaussian fit of the ΔE signal for (200, 3000) range is shown in the Fig.5b.

Therefore, for the case a) **the** weight function $W_i$ will be as

$$W_i(TOF_i, \Delta E_i, E_i^{REG}, t_i) = F(TOF_i, \Delta E_i) \cdot (1-e^{-\frac{t_i}{\tau}}) \cdot \frac{1}{\sigma_i\sqrt{2\pi}} \cdot e^{-\frac{\Delta\varepsilon_i^2}{2\sigma_i^2}}\ (6),$$

where $the\ E_i^{REG}$ value is calculated from (4) and $\sigma_i$ is standard deviation, usually of about 2 MeV, and $\Delta\varepsilon_i = E_i^{REG} - <E^{REG}>.$

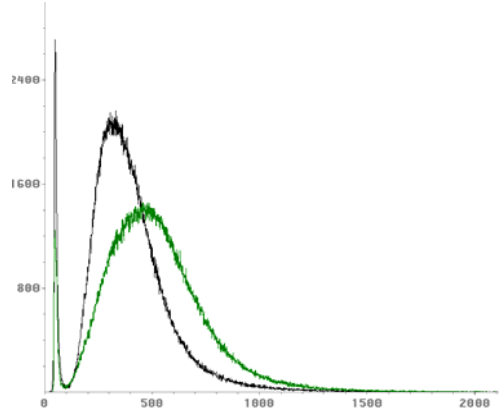

Fig.5a Typical shapes for EVR ΔE signals (channels) measured with START and STOP proportional chambers. The pentane pressure value is equal to 1.6 Torr. Both anode biases are equal to +390 Volts. Both cathode biases are equal to -100 Volts. Reaction: $^{nat}Yb+^{48}Ca \rightarrow\ ^*Th$.

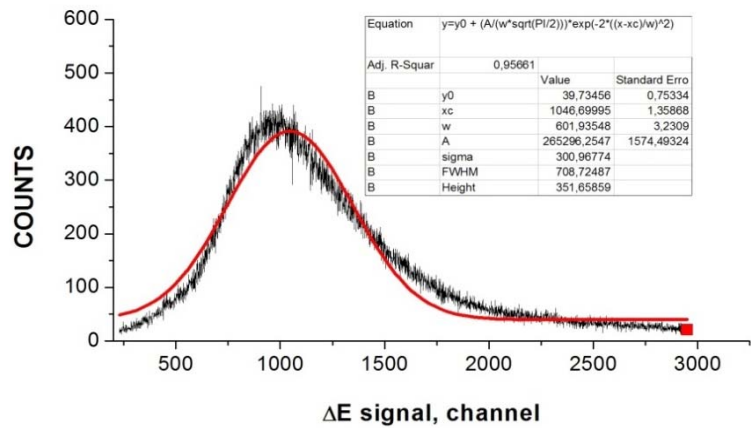

Fig.5b Gaussian fit of the ΔE signal for (200, 3000) range. Fit parameters are within the frame

d) The case of few candidates to α-decay is not significant enough to be taken into consideration due to a much lower rate of signals simulating alpha decay in the focal plane detector of the DGFRS.

4. **Example of equation (1) solution for n=2**

In Figure 6 (a,b), a ten-step simple iteration process for the parameter $\Delta = \frac{t_2 - t_1}{\tau} = 0.2$ and the initial approximation $x_0 = 0.3$ are shown. Note that although the total execution time is equal to **6 μS**, the 1.2-μS interval (two steps) meets the requirements fulfils the conditions of perfect application of the real-time technique. Note that Newton method gives nearly the same convergence time in comparison with the simple iteration one.

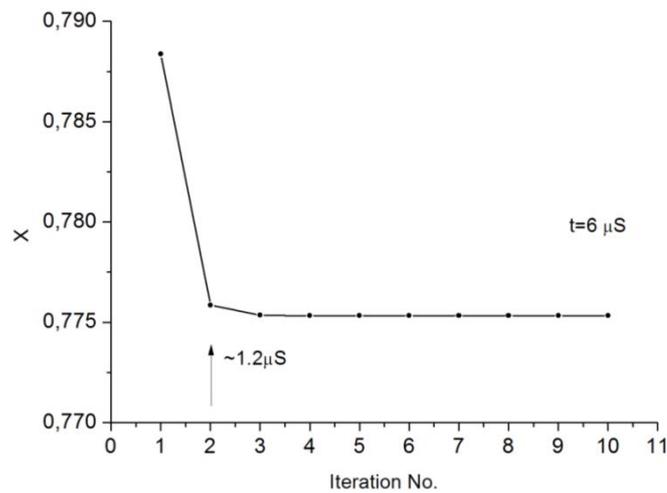

Fig.6a. Solution of equation (1) against the number of simple iterations. $X = t_1/\tau$.

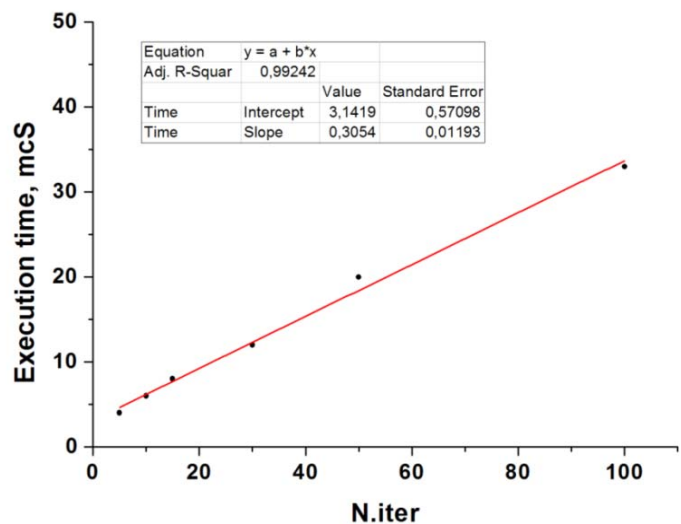

Fig.6b. Dependence of execution time against iteration number

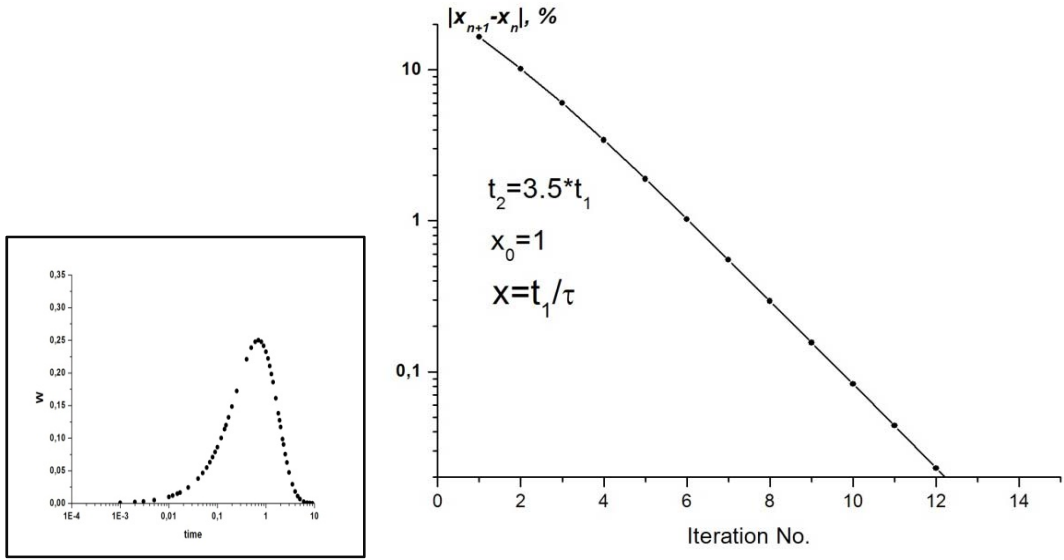

Fig.6c Process of convergence (right side) of sequence $s_n=|x_{n+1}-x_n|$ for a statistical weight of

$W_i \approx e^{-\frac{t_i}{\tau}} \cdot \left(1 - e^{-\frac{t_i}{\tau}}\right)$ $n = 2$ **(left side)**. Note, that very satisfied results in the sequence $s_n$ convergence are achieved with $\tau_0 \approx \sqrt[2]{t_1 t_2}$ value is taken as a first approximation.

## 5. Application of the combined method

The author has not excluded the combined (relatively recoil signal) method application.

In other words, when detecting a recoil signal, a shorter beam-off interval is generated by the DGFRS detection system. This trivial algorithm could obviously operate in parallel with the main one ("***OR***" principle), like it is shown in Fig.7.

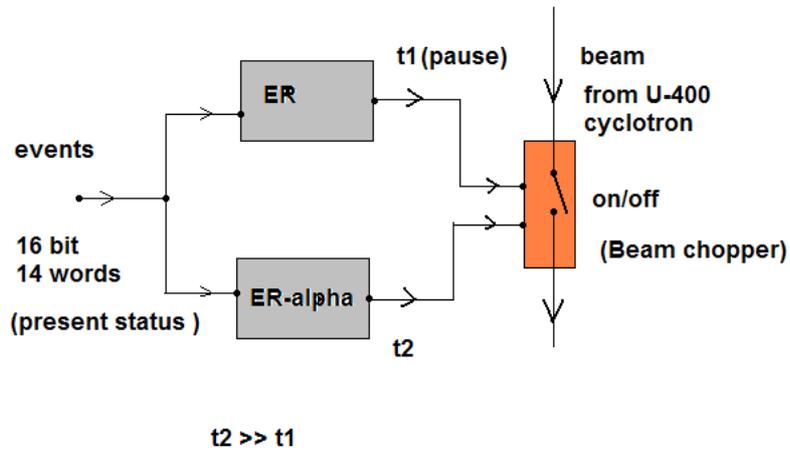

**Fig.7. Schematics of the combined beam-stop real-time algorithm.**

## 6. Summary

Together with conservative approaches minimizing beam associated backgrounds, such as the construction of new electromagnetic recoil separators, fast chemistry, and design of a more perfect silicon radiation detector, the development of the "active correlation" method will definitely contribute to this aim. Moreover, this method will provide radical suppression of background products. The extension of the method is presented above and will undergo exhaustive beam tests in the very near future.

Of course, the development of new gas-filled recoil separators will contribute to the problem of background suppression too.

The author thanks Drs. A. Kuznetsov and A.Voinov for their help in some test measurements for this paper. The paper is supported in part by the RFBR Grant No.13-02-12052.

**Supplement 1. A few words about basic electronics module implementation**

The first approach to the DGFRS spectrometer implementation was discussed in detail in [8]. This decision is based on single 16 in 12-bit ADC PA-1 about 1 μS conversion time application [8] and a special digital unit used to obtain the address of the back strip number of the DSSSD detector. Consequently, the description of the C++ code reported in [8] is outside the scope of the present paper. Another reasonable scenario is based on the application of the universal CAMAC 1M 16 amplifier-multiplexer-ADC integrated module ADP16 produced by *TekhInvest Dubna* [16]. Below the author is reporting the results of a preliminary test of the first TekhInvest module using the IMI2011 special purpose generator module[v] [14]. To carry out the test, the author designed the Builder C++ (Windows) code. The mentioned module has **a** capability to store eight signal amplitudes together with their synchronized times in the internal buffer memory. It allows detection of short (~2.5 μS) sequences of signals. Therefore, the signal sequences of $x_1 \rightarrow x_2 \rightarrow ... x_n : n \leq 8: 0 \rightarrow 2.5\mu S \rightarrow ... \rightarrow 2.5 \rightarrow n * \tau_{DEAD}$ ( writing time – reading time) will be successfully detected with their time stamps with the microsecond accuracy. Here, $\tau_{dead}$ is the regular dead time value of the spectrometer. The $\tau_{dead}$ value depends on the actual spectrometer crate configuration (the number of actual stations and their CAMAC functions in use). The tested module has the following CAMAC functions:

- N*A(0)[F(0)+F(2)] alpha-particle scale reading (13 bit),
- N*A(1)[F(0)+F(2)] fission fragment (heavy-ion) scale reading(12 bit),
- N*A(2)[F(0)+F(2)] synchronized time (in μS) reading,

- N*A(0)*F(8)      test LAM,
- N*A(0)*F(10)     data reset,
- N*A(0)*F(16)   W(8..1) threshold setting,
- N*A(0)*F(24)   masking L, and
- N*A(0)*F(26)   de-masking L.

In Fig.8a,b test spectra are shown in the right part of main window. FF-scale spectrum is shown in the left upper corner. Additionally, one general conclusion can be drawn here.

*The prototype of new Builder C++ based software for the DGFRS has been successfully tested.*

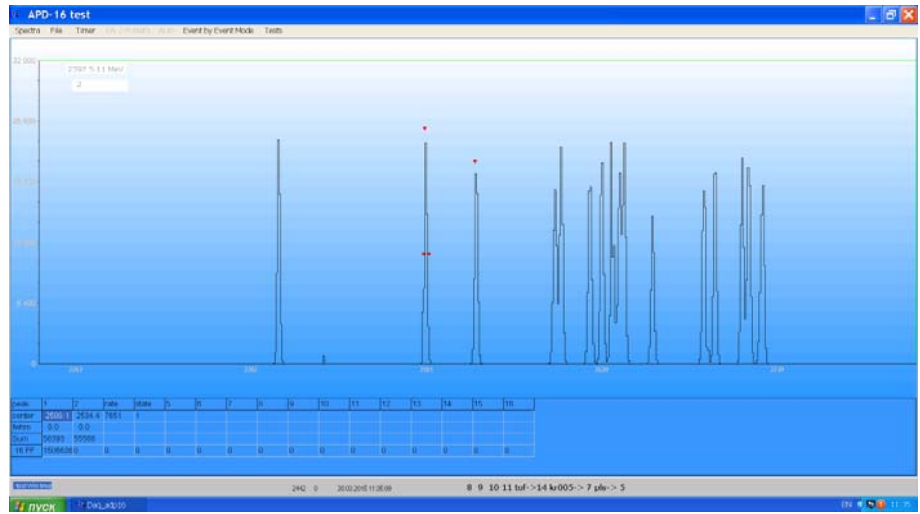

**Fig.8a** Test spectra from IMI2011-ADP16. The bottom line of the table in the figure shows event number per each channel. Peaks 1,2 under the stability test are shown by the triangles.

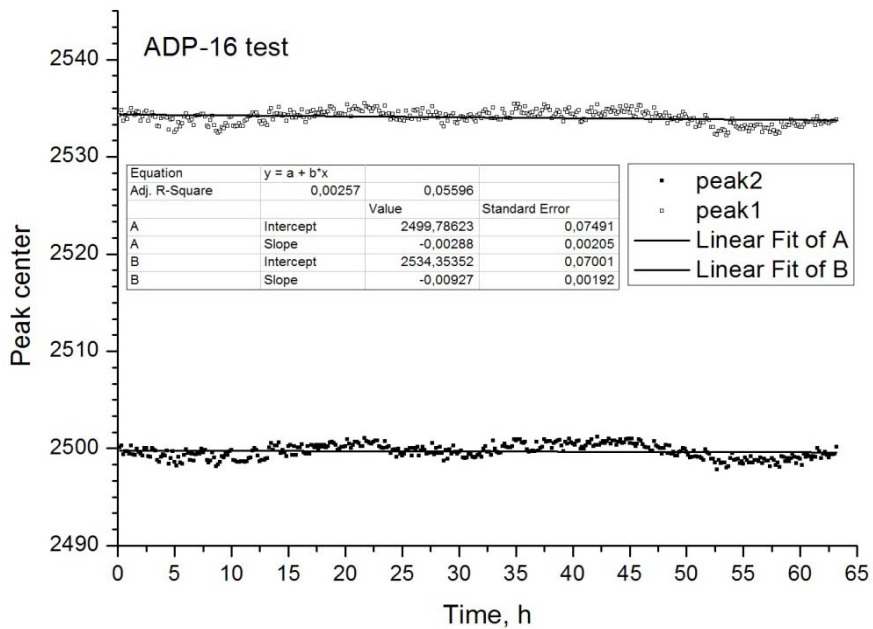

Fig. 8b Stability test for ADP-16(left peak in the 6a). Linear fit results are presented in the table within.

The standard deviation value is equal to 0.07 channels for both peaks. With using special design thermo stabilized resistors the whole stability will be more perfect.

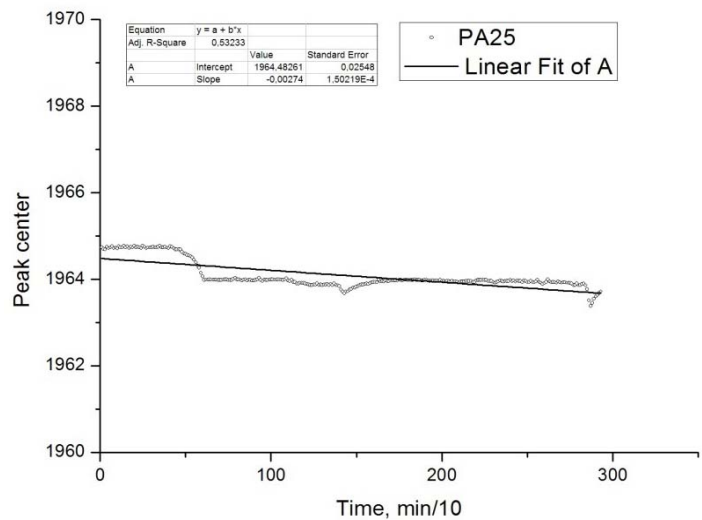

Fig.8c Stability test for ADC PA-25. Standard deviation is equals to 0.025 channels.

## 7. References

[1] OganessianYu.Ts., Utyonkov V.K. Superheavy element research. //Rep. Prog. Phys. 2015 (in print)

[2] TsyganovYu.S., Polyakov A.N., Sukhov A.N. An improved real-time PC based algorithm for extraction of recoil-alpha sequences in heavy-ion induced nuclear reactions.// Nucl. Instr. Meth. A. 2003. Vol.513. p.413-416

[3] TsyganovYu.S., Polyakov A.N. Real-time operating mode with DSSSD detector to search for short correlation ER-alpha chains. // Cybernetics and Phys. 2014. Vol.3, No.2.p.85-90

[4] TsyganovYu.S. Method of "active correlations" for DSSSD detector application. // Phys. Part. and. Nucl.Lett. 2015. Vol.12, No.1. p.83-88.

[5] OganessianYu.Ts., Utyonkov V.K, LobanovYu.V., AbdullinF.Sh., Polyakov A.N., Shirokovsky I.V., Tsyganov Yu.S.,Gulbekyan G.G., Bogomolov S.L.,Gikal B.N., Mezentsev A.N., Iliev S.,Subbotin V.G., Sukhov A.M.,Buklanov G.V., Subotic K., Itkis M.G., Moody K.J., Wiln J.F., Stoyer N.J., Stoyer M.A., Lougheed R.W. The synthesis of

superheavy nuclei in the $^{48}$Ca+$^{244}$Pu reaction. // Revista Mexicana de Fisica. 2000. Vol.46. supplemento 1. p.35-41

[6] LazarevYu.A., LobanovYu.V., OganessianYu.Ts., Utyonkov V.K., AbdullinF.Sh., Polyakov A.N., Rigol J., Shirokovsky I.V., TsyganovYu.S., Iliev S., Subbotin V.G., Sukhov A.M., Buklanov G.V.,Gikal B.N., Kutner V.B., Mezentsev A.N., Subotic K., Wild J.F.,LougheedR.W.,Moody K.J. α-decay of $^{273}$110: Shell closure at N=162. // Phys.

[7] Zlokazov V.B, TsyganovYu.S. Half-life estimation under indefinite "mother-daughter" relation. // Phys. Part. Nucl.Lett. 2010. Vol.7, No.6. p.401-405

[8] TsyganovYu.S. Elemennts of automation of the DGFRS experiments. // Lett. To ECHAYA. 2015. Vol.12, No.1(192) pp.116-127

[9] TsyganovYu.S. Synthesis of new superheavy elements at the DGFRS: complex of technologies. ECHAYA 2014.Vol.45 No.5-6. P.1514

[10] TsyganovYu.S. Parameter of equilibrium charge states distribution width for calculation of heavy recoil spectra.// Nucl. Inst. Meth. A. 1996. Vol.378. p.356-359

[11] Mezentsev A.N., Polyakov A.N., TsyganovYu.S.,Subbotin V.G., Ivanova I. Low pressure TOF module.// FLNR (JINR) Sci. Report 1992-1993.Dubna 1993. P.203

[12] TsyganovYu.S., Subbotin V.G., Polyakov A.N., Sukhov A.M., Iliev S., Mezentsev A.N., Vacatov D.V. Focal plane detector of the Dubna gas-filled recoil separator.// Nucl. Instr. Meth. A. 1997. Vol.392. p.197-201

[13] Tsyganov Yu.S. A new reasonable scenario to search for ER-alpha energy-time-position correlated sequences in a real time mode. // Lett. To ECHAYA. 2015. Vol.12, No.4

[14] ADP-16 1M module, IMI2011 module. // Technical manual of "TechInvest" (free economy zone Dubna)

---

[i] For the sake of the statistical significance
[ii] E.g. ε=0.5
[iii] An alternative, simpler, totally empirical dependence is: $E_{REG} \approx$ -1.7+0.74·$E_{in}$.
[iv] If necessary, one can take into account more exact approximation, e.g. Poisson like function.
[v] The module generates 16 channels of signals similar to ones from a spectroscopy shape amplifier